\documentclass[conference]{IEEEtran}
\IEEEoverridecommandlockouts
\usepackage{cite}
\usepackage{amsmath,amssymb,amsfonts}
\usepackage{algorithmic}
\usepackage[dvipdfmx]{graphicx}
\usepackage{textcomp}
\usepackage{xcolor}
\usepackage{mathrsfs}
\usepackage{multirow}
\usepackage{comment}
\usepackage{here}
\usepackage{cases}
\usepackage{empheq}
\usepackage{arydshln}
\usepackage{bm}
\newcommand{\her}{\mathsf{H}}
\newcommand{\trp}{\mathsf{T}}
\newcommand{\subt}{\mathrm{(s)}}
\newcommand{\subn}{\mathrm{(n)}}
\newcommand{\subtn}{{(l)}}
\newcommand{\cG}[2]{\mathcal{N}_\mathbb{C}\left(#1,#2\right)}
\allowdisplaybreaks
\usepackage{authblk}
\usepackage {subfig}
\renewcommand{\baselinestretch}{0.975}
\def\BibTeX{{\rm B\kern-.05em{\sc i\kern-.025em b}\kern-.08em
    T\kern-.1667em\lower.7ex\hbox{E}\kern-.125emX}}
\begin{document}

\title{NoisyILRMA: Diffuse-Noise-Aware \\Independent Low-Rank Matrix Analysis for \\Fast Blind Source Extraction \\
\thanks{This research was partly supported by JST Moonshot R\&D Grant Number JPMJMS2011 and JSPS KAKENHI Grant Number 19H01116.}
}
\renewcommand\Affilfont{\normalsize}
\setlength{\affilsep}{0em}
\author[$\dag$]{Koki Nishida}
\author[$\dag$]{Norihiro Takamune}
\author[$\ddag$]{Rintaro Ikeshita}
\author[$\S$]{Daichi Kitamura}
\author[$\dag$]{\authorcr Hiroshi Saruwatari}
\author[$\ddag$]{Tomohiro Nakatani}
\affil[$\dag$]{The University of Tokyo, Graduate School of Information Science and Technology, Tokyo, Japan}
\affil[$\ddag$]{NTT Communication Science Laboratories, NTT Corporation, Kyoto, Japan}
\affil[$\S$]{National Institute of Technology, Kagawa College, Kagawa, Japan}

\maketitle

\begin{abstract}
In this paper, we address the multichannel blind source extraction (BSE) of a single source in diffuse noise environments. To solve this problem even faster than by fast multichannel nonnegative
matrix factorization (FastMNMF) and its variant, we propose a BSE method called NoisyILRMA, which is a modification of independent low-rank matrix analysis (ILRMA) to account for diffuse noise.
NoisyILRMA can achieve considerably fast BSE by incorporating an algorithm developed for independent vector extraction. In addition, to improve the BSE performance of NoisyILRMA, we propose a mechanism to switch the source model with ILRMA-like nonnegative matrix factorization to a more expressive source model during optimization. In the experiment, we show that NoisyILRMA runs faster than a FastMNMF algorithm while maintaining the BSE performance. We also confirm that the switching mechanism improves the BSE performance of NoisyILRMA.
\end{abstract}


\section{Introduction}
Multichannel blind source separation (BSS) is a technique used to separate multiple sources from multichannel observed signals recorded by a microphone array without any prior knowledge of, for example, the characteristics of sources or spatial mixing systems.  Among BSS, the technique used to extract source signals from the background noise is particularly called multichannel blind source extraction (BSE). BSE can be used as a front-end of sound signal processing devices such as hearing aids and smart speakers. This study is particularly focused on extracting one target source in diffuse noise environments.


\begin{figure}[tbp]
\centering
\includegraphics[width=8.7cm]{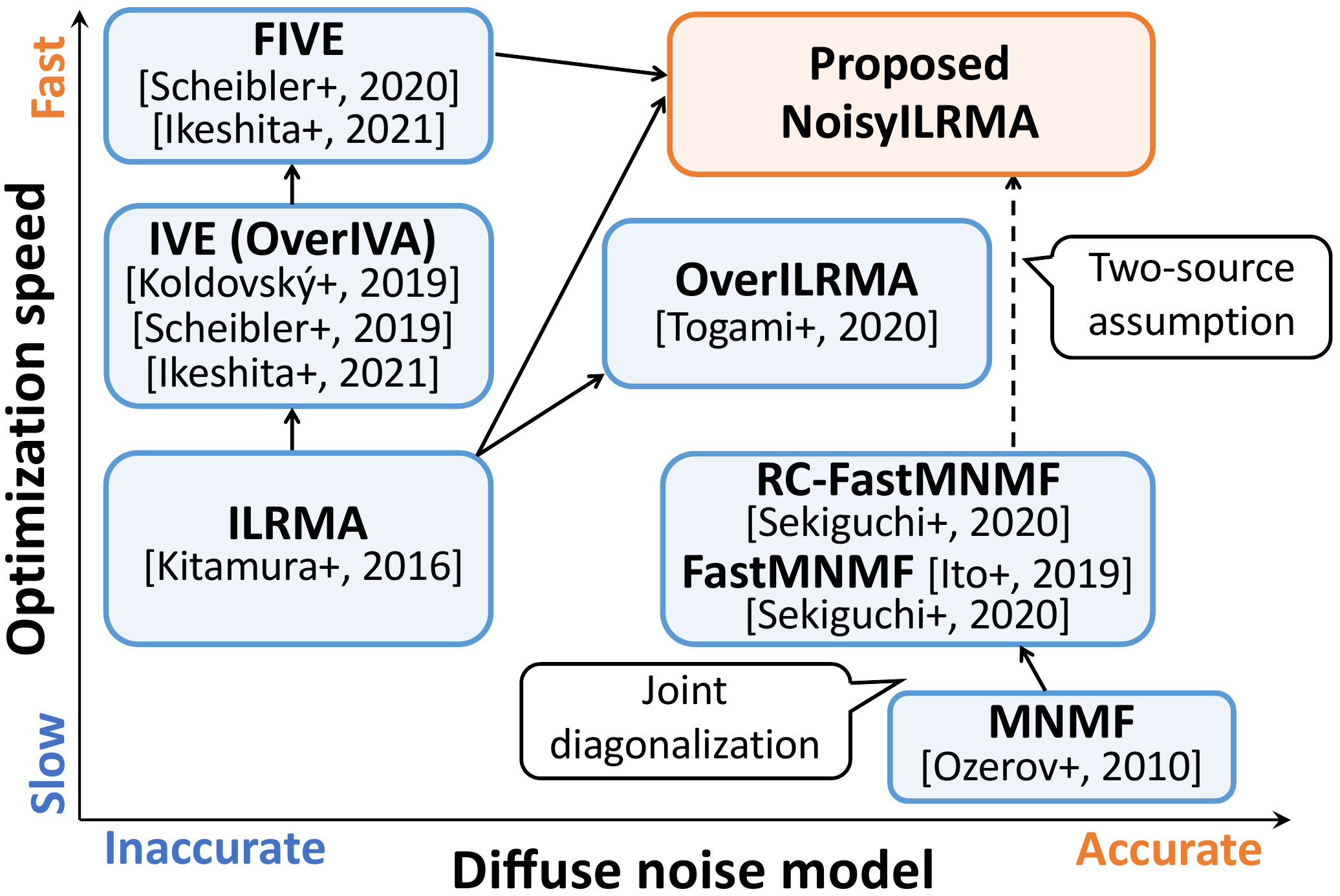}
\caption{Relationships between proposed and conventional methods.}
\label{fig:relation}
\end{figure}

Independent low-rank matrix analysis (ILRMA)~\cite{Kitamura2016} is one of the BSS methods that can separate point sources when the number of sources is less than or equal to that of microphones. 
ILRMA assumes independence between the sources and the low rankness of the sources in the time-frequency domain using nonnegative matrix factorization (NMF)~\cite{NMF}.
In \cite{Kitamura2016}, ILRMA was reported to experimentally achieve high and stable performance.
Although this method can separate point sources accurately, it is impossible to remove diffuse noise in the same direction as the target source~\cite{Araki2003}.

Multichannel NMF (MNMF)~\cite{Ozerov2010} is a multichannel extension of NMF employing a full-rank spatial covariance matrix (SCM)~\cite{Duong2010} for each source, which can model diffuse noise. However, its computational complexity is high owing to the large number of parameters for full-rank SCMs. To overcome this problem, FastMNMF~\cite{Ito2019, Sekiguchi2020} has been proposed. FastMNMF assumes that the SCMs for all sources are jointly diagonalizable~\cite{JD}, which allows a faster algorithm called iterative projection (IP)~\cite{IP} to be used for the estimation of the joint-diagonalization matrix.
By restricting the situation where some sources are point sources, rank-constrained FastMNMF (RC-FastMNMF) ~\cite{Sekiguchi2020} has been proposed for efficient estimation. This method focuses on the fact that the SCM of a point source can be approximated by a rank-1 matrix and constrains the SCMs corresponding to the point sources as rank-1 matrices, enabling the efficient exploration of the solution space.

In this paper, we propose a method called NoisyILRMA, which is ILRMA modified to account for diffuse noise. When we use ILRMA under diffuse noise environments, the separated signal corresponding to the target source contains the target source and noise, while the other separated signals contain noise only~\cite{Takahashi2009}. 
By explicitly modeling this property, NoisyILRMA simultaneously estimates the separated target signal and the residual noise component in it.
We focus on the fact that the optimization problem of the demixing filters has the same form as that in independent vector extraction (IVE)~\cite{IVE,OverIVA,Ikeshita2021}, which is a method used to efficiently extract the target sources in overdetermined cases (the number of sources is less than that of microphones). 
For IVE, fast IVE (FIVE)~\cite{FIVE,Ikeshita2021}, in which the demixing filters are optimized by a fast algorithm, has been proposed.
 From these facts, we can use the same fast algorithm as FIVE 
 to optimize the demixing filters in NoisyILRMA.
 In Sect.~\ref{sec:relation}, we show that NoisyILRMA can be viewed as an accelerated RC-FastMNMF used under the two-source assumption (for single point source extraction under diffuse noise). We also discuss OverILRMA~\cite{OverILRMA}, which is an extension of ILRMA to overdetermined cases, as another closely related method in Sect.~\ref{sec:relation}.
Fig.~\ref{fig:relation} shows the relationships between NoisyILRMA and other methods.

In addition, to further improve the BSE performance of NoisyILRMA, we propose a source model switching mechanism. 
In NoisyILRMA, the limited expressiveness of the NMF model degrades the final BSE performance, although the NMF model is useful for estimating the demixing filters.
To solve this problem, we focus on rank-constrained SCM estimation (RCSCME)~\cite{Kubo2020}. By utilizing the spatial information estimated in the preprocessing BSS method such as ILRMA, RCSCME efficiently estimates the residual spatial information and the source models for the target source and noise. RCSCME achieves high-performance BSE owing to the use of a more expressive source model than the NMF model.
We focus on this fact and propose a source model switching mechanism in which we first obtain the demixing filters accurately using the NMF model and then fix the demixing filters and switch to the same highly expressive source model as in RCSCME. 


\section{Background}
\subsection{ILRMA~\cite{Kitamura2016}} \label{sec:ILRMA}
Let $\bm{x}_{ij}\in\mathbb{C}^M$ be the short-time Fourier transformation (STFT) of the multichannel observed signal, where $i$ and $j$ are the indices of frequency bins and time frames, respectively, and $M$ is the number of microphones. When each source can be assumed to be a point source and the window in STFT is sufficiently longer than the reverberation, the following mixing system
$
	\bm{x}_{ij} = \mathbf{A}_i\bm{s}_{ij} \label{eq:xILRMA}
$
holds, 
where $\bm{s}_{ij}\in\mathbb{C}^N$ and $\mathbf{A}_i = (\bm{a}_{i1},\dots,\bm{a}_{iN})$ are the source signals and mixing matrix, respectively. Here, $\bm{a}_{in}\in\mathbb{C}^M$ denotes the steering vector, where $n\in\{1,\dots,N\}$ is the index of the sources. If $N = M$ and $\mathbf{A}_i$ is invertible, the separated signal $\bm{y}_{ij} = (y_{ij1},\dots,y_{ijM})^\trp$ can be obtained as
\begin{align}
	\bm{y}_{ij} = \mathbf{W}_i^\her\bm{x}_{ij},\label{eq:y=Wx}
\end{align}
where $\mathbf{W}_i = (\mathbf{A}_i^\her)^{-1}$ is the demixing matrix, and $\mbox{}^\trp$ and $\mbox{}^\her$ denote transpose and Hermitian transpose, respectively.
We assume  the following complex Gaussian distribution for $y_{ijn}$:
\begin{align}
	y_{ijn}\sim\cG{0}{r_{ijn}},
\end{align}
where 
$r_{ijn}$ is the time-variant variance of source $n$, which is modeled by NMF as $r_{ijn} = \sum_k t_{ikn} v_{kjn}$. Here, $t_{ikn}, v_{kjn}\geq 0$ are the NMF basis and activation, respectively, and $k \in \{1,\dots,K\}$ is the index of the basis. 

When there exist a single point source and diffuse noise, the separated signals satisfy the following property~\cite{Takahashi2009}:
$y_{ijn_\mathrm{s}}$ contains both the target source and noise, while the other $M-1$ separated signals contain noise only,
where $n_\mathrm{s}$ is the index of the separated signal corresponding to the target source.

\subsection{RC-FastMNMF~\cite{Sekiguchi2020}}
In MNMF, we assume the following multivariate complex Gaussian distribution for $\bm{x}_{ij}$:
\begin{align}
	\bm{x}_{ij} \sim \mathcal{N}_\mathbb{C}\Biggl(\bm{0}_M, \sum_{\tilde{n}=1}^{\tilde{N}}r_{ij\tilde{n}}\mathbf{R}_{i\tilde{n}}\Biggr),\label{eq:GenRC}
\end{align}
where $\bm{0}_M\in\mathbb{C}^M$ is a zero vector, $\mathbf{R}_{i\tilde{n}}\in\mathbb{C}^{M\times M}$ denotes the SCM of source $\tilde{n}$, and $\tilde{n}\in\{1,\dots,\tilde{N}\}$ is the index of sources. Here, $r_{ij\tilde{n}}$ is also modeled by NMF as $r_{ij\tilde{n}} = \sum_k t_{ik\tilde{n}} v_{kj\tilde{n}}$.
Note that $\tilde{N}$ is not necessarily equal to $M$, unlike in ILRMA.
 In FastMNMF, we assume the following joint diagonalizability for the SCMs of the $\tilde{N}$ sources to estimate them efficiently:
\begin{align}
	\tilde{\mathbf{W}}_i^\her\mathbf{R}_{i\tilde{n}}\tilde{\mathbf{W}}_i &= \mathrm{diag}\left(\lambda_{i1\tilde{n}},\dots,\lambda_{iM\tilde{n}}\right),\label{eq:JD}
\end{align}
where $\mathrm{diag}(q_1\dots,q_M)\in\mathbb{C}^{M\times M}$ is the diagonal matrix whose $m$th element is $q_m$, $\tilde{\mathbf{W}}_i = (\tilde{\bm{w}}_{i,1},\dots,\tilde{\bm{w}}_{iM})\in\mathbb{C}^{M\times M}$ is the joint-diagonalization matrix, $\lambda_{im\tilde{n}}\geq0$ is a diagonal element of diagonalized SCMs, and $m\in\{1,\dots,M\}$ is the index of the column of the joint-diagonalization matrix. 
In RC-FastMNMF, from the fact that the SCM of a point source can be approximated as a rank-1 matrix,  we introduce the rank-1 constraint for the point source $\tilde{n}'$ by setting $\lambda_{im\tilde{n}'} = 0$ for $m\in\{1,\dots,M\}\backslash\{\tilde{n}'\}$.

The model parameters of RC-FastMNMF $\Theta^{\mbox{\footnotesize RC-FastMNMF}} = \{t_{ik\tilde{n}},v_{kj\tilde{n}}, \tilde{\mathbf{W}}_i, \lambda_{im\tilde{n}}\}$ are estimated in the maximum likelihood sense. For $\tilde{\mathbf{W}}_i$, the IP algorithm~\cite{IP} can be used~\cite{Sekiguchi2020}, where each $\tilde{\mathbf{W}}_i$ column is alternately updated. A relationship with the proposed method is discussed in Sect.~\ref{sec:relation}.
\subsection{RCSCME~\cite{Kubo2020}}
Using the property of ILRMA under diffuse noise environments, we can accurately obtain the steering vector of the target source $\bm{a}'_i\in\mathbb{C}^M$ and the rank-$(M-1)$ component of the noise SCM $\mathbf{R'}_i^\subn\in\mathbb{C}^{M\times M}$. Focusing on this fact, RCSCME uses ILRMA as a preprocess and utilizes the spatial information obtained in ILRMA for efficient estimation.

In RCSCME, we assume $\bm{x}_{ij}$ follows the following multivariate complex Gaussian distribution with the inverse-gamma prior distribution:
\begin{align}
	 \bm{x}_{ij}\mid r_{ij}^\subt &\sim \mathcal{N}_\mathbb{C} \left(\bm{0}_M,r_{ij}^\subt\bm{a}'_i(\bm{a}'_i)^\her + r_{ij}^\subn(\mathbf{R'}_i^\subn + \mu_i \bm{b}_i\bm{b}_i^\her)\right),\nonumber\\
	 r_{ij}^\subt &\sim \mathcal{IG}(\alpha,\beta),\label{eq:RCSCME_prior}
\end{align}
where $r_{ij}^\subt, r_{ij}^\subn > 0$ are time-variant variances of the target speech and noise, $\mu_i>0$ and $\bm{b}_i\in\mathbb{C}^M$ are the weight and direction vector used to represent the deficient rank-1 component of the noise SCM, and $\alpha,\beta>0$ are the shape and scale parameters of the inverse-gamma distribution, respectively.
Here, $\bm{a}'_i = (\mathbf{W'}_i^\her)^{-1}\bm{e}_{n_\mathrm{s}}$ holds, where $\mathbf{W}'_i$ is the demixing matrix estimated in ILRMA and $\bm{e}_n$ is the unit vector whose $n$th element is one. 
We calculate $\mathbf{R'}_i^\subn$ as $\mathbf{R'}_i^\subn = \sum_j{\bm{x}'}_{ij}^\subn({\bm{x}'}_{ij}^\subn)^\her/J$, where ${\bm{x}'}_{ij}^\subn = \bm{x}_{ij}-\bm{a}'_i(\mathbf{W}'_i\bm{e}_{n_\mathrm{s}})^\her\bm{x}_{ij}$ holds, and $J$ is the number of time frames. $\bm{b}_i$ is a constant vector that is linearly independent of the column vectors of $\mathbf{R'}_i^\subn$. The inverse-gamma prior distribution in (\ref{eq:RCSCME_prior}) is introduced for the sparsity of the target source in the time-frequency domain.
Note that $r_{ij}^\subt$ and  $r_{ij}^\subn$ are unconstrained parameters with the prior distribution in (\ref{eq:RCSCME_prior}), which is more expressive than the NMF model.
RCSCME can achieve high BSE performance owing to the more expressive source model.
\section{Proposed methods}\label{sec:propose}
\subsection{Motivation}
In this section, we propose the diffuse-noise-aware ILRMA, which we call NoisyILRMA, for single-source extraction. 
When we use ILRMA in diffuse noise environments, the following properties hold~\cite{Takahashi2009}:
\begin{itemize}
\item $y_{ijn_\mathrm{s}}$ contains both speech and noise.
\item The other $M-1$ separated signals contain noise only.
\end{itemize}
By explicitly modeling these properties, NoisyILRMA simultaneously estimates the demixing matrix $\mathbf{W}_i$ and the noise component in the separated target signal $y_{ijn_\mathrm{s}}$.
In NoisyILRMA, we also assume that the noise component in $y_{ijn_\mathrm{s}}$ has time synchronization with the other $M-1$ separated signals to enable a reasonable estimation.
We can suppress the noise component in $y_{ijn_\mathrm{s}}$ by multichannel Wiener filter (MWF) using the estimated variances.
As an additional advantage of such modeling in NoisyILRMA, we show that the optimization problem of the demixing matrix has the same form as that in IVE\cite{IVE,OverIVA,Ikeshita2021},
which enables us to use the same considerably fast algorithm in FIVE~\cite{FIVE,Ikeshita2021}.

In addition, to further improve the BSE performance of NoisyILRMA, we propose a source model switching mechanism. 
The NMF model in NoisyILRMA is useful for estimating the demixing matrix $\mathbf{W}_i$ because it clusters the frequency bins corresponding to the same source. However, the NMF model may degrade the final BSE performance when using the MWF owing to its limited expressiveness. Inspired by RCSCME, in the proposed switching mechanism, we first obtain $\mathbf{W}_i$ accurately using the NMF model and then switch to the same highly expressive source model as in RCSCME.
\subsection{Method}
In NoisyILRMA, we introduce the above-mentioned properties of the separated signal $y_{ijn}$ and the assumption that the noise component in $y_{ijn_{\mathrm{s}}}$ has time synchronization with the other separated signals as follows:
\begin{align}
	y_{ij1} &\sim \mathcal{N}_\mathbb{C}\left(0, r_{ij}^\subt + r_{ij}^\subn \lambda_i^\subn\right)\label{eq:RC_yt},\\
	y_{ijn} &\sim \mathcal{N}_\mathbb{C}\left(0, r_{ij}^\subn\right) ~(n \in\{ 2,\dots, M\})\label{eq:RC_yn},
\end{align}
where $\lambda_i^\subn > 0$ denotes the weight of the noise component in $y_{ij1}$, $r_{ij}^\subtn$ is modeled by NMF as $r_{ij}^\subtn=\sum_kt_{ik}^\subtn v_{kj}^\subtn$, $l\in\{\mathrm{s,n}\}$ is the label used to distinguish the target source and noise, and $t_{ik}^\subtn$, $v_{kj}^\subtn\geq0$ are the NMF basis and activation, respectively.
Note that we assume $n_\mathrm{s} = 1$ 
without loss of generality and $y_{ij2},\dots,y_{ijM}$
  to have the same variance by using the scale arbitrariness of $\bm{w}_{in}$.

In NoisyILRMA, 
the cost function is defined as the negative log-likelihood, which is obtained from (\ref{eq:y=Wx}), (\ref{eq:RC_yt}), and (\ref{eq:RC_yn}) as
\begin{align}
	\mathcal{L}(\Theta) &= \sum_{i,j} \Biggl[-2\log|\det{\mathbf{W}_i}|  \nonumber\\
	&+ \log{(r_{ij}^\subt + r_{ij}^\subn\lambda_i^\subn)} +(M-1)\log{r_{ij}^\subn} \nonumber\\
	 &+ \frac{|y_{ij1}|^2}{r_{ij}^\subt + r_{ij}^\subn\lambda_i^\subn} + \frac{\sum_{n=2}^M|y_{ijn}|^2}{r_{ij}^\subn} \Biggr] +\mathrm{const.},\label{eq:cost_NoisyILRMA}
\end{align}
where $\Theta = \{t_{ik}^\subtn,v_{kj}^\subtn,\mathbf{W}_i,\lambda_i^\subn\}$ is the set of model parameters and $\mathrm{const.}$ includes the terms independent of $\Theta$. 
 $\mathbf{W}_i$ and the other parameters $\{t_{ik}^\subtn,v_{kj}^\subtn,\lambda_i^\subn\}$ are alternately updated to minimize (\ref{eq:cost_NoisyILRMA}).
 To derive the update rule for $\mathbf{W}_i$,
we transform the cost function (\ref{eq:cost_NoisyILRMA}) with respect to $\mathbf{W}_i = (\bm{w}_i^\subt,\mathbf{W}_i^\subn)$ as
\begin{align}
	    \mathcal{L}(\{\mathbf{W}_i\}) 
        &= J\sum_i \biggl[- 2 \log |\det \mathbf{W}_i| + (\bm{w}_i^\subt)^\her \mathbf{G}_i^\subt \bm{w}_i^\subt 
        \nonumber\\
        &\quad 
        + \mathrm{Tr} \Big(
            (\mathbf{W}_i^\subn)^\her \mathbf{G}_i^\subn \mathbf{W}_i^\subn
        \Big) \biggr] + \mathrm{const.},\label{eq:cost_NoisyILRMA_W}
\end{align}
where we define $\mathbf{G}_i^\subt = \sum_j \bm{x}_{ij}\bm{x}_{ij}^\her/(r_{ij}^\subt+ r_{ij}^\subn \lambda_i^\subn)/J$ and $\mathbf{G}_i^\subn = \sum_j \bm{x}_{ij}\bm{x}_{ij}^\her/r_{ij}^\subn/J$, and $\mathrm{const.}$ includes the terms independent of $\mathbf{W}_i$.
Since this cost function (\ref{eq:cost_NoisyILRMA_W}) is the same form as that for IVE~\cite{IVE,OverIVA,Ikeshita2021}, the fast algorithm in FIVE~\cite{FIVE,Ikeshita2021} can be used to optimize $\mathbf{W}_i$.
By transforming the condition that the Wirtinger derivative of (\ref{eq:cost_NoisyILRMA_W}) with respect to $\mathbf{W}_i$ equals zero, we can obtain the following equations:
\begin{align}
	(\bm{w}_i^\subt)^\her\mathbf{G}_i^\subt\bm{w}_i^\subt &= 1,\label{eq:wKtw}\\
	(\mathbf{W}_i^\subn)^\her\mathbf{G}_i^\subt\bm{w}_i^\subt &= \bm{0}_{M-1},\label{eq:wKtwn}\\
	(\bm{w}_i^\subt)^\her\mathbf{G}_i^\subn\mathbf{W}_i^\subn &= \bm{0}_{M-1}^\her,\label{eq:wKnwn}\\
	(\mathbf{W}_i^\subn)^\her\mathbf{G}_i^\subn\mathbf{W}_i^\subn &= \mathbf{E}_{M-1},\label{eq:wnKnwn}
\end{align}
where $\mathbf{E}_{M-1}\in\mathbb{C}^{(M-1)\times(M-1)}$ is the identity matrix. In \cite{Ikeshita2021}, the update rule of $\bm{w}_i^\subt$ is derived as
\begin{align}
	\bm{w}_i^\subt &\leftarrow \frac{\bm{h}_{i1}}{\sqrt{\bm{h}_{i1}^\her\mathbf{G}_i^\subt\bm{h}_{i1}}},\label{eq:GEP_wt}
\end{align}
where $\bm{h}_{i1}\in\mathbb{C}^M$ is the generalized eigenvector with the largest generalized eigenvalue in the following generalized eigenvalue problem:
\begin{align}
	\mathbf{G}_i^\subn\bm{v}_i = \kappa_i~ \mathbf{G}_i^\subt\bm{v}_i.\label{eq:GEP}
\end{align}
Here, $\kappa_i > 0$ and $\bm{v}_i\in\mathbb{C}^M$ are the generalized eigenvalue and the generalized eigenvector, respectively. We can update $\mathbf{W}_i^\subn$ to satisfy (\ref{eq:wKtwn})--(\ref{eq:wnKnwn}) as follows:
\begin{align}
	\mathbf{W}_i^\subn &\leftarrow \left(\frac{\bm{h}_{i2}}{\sqrt{\bm{h}_{i2}^\her\mathbf{G}_i^\subn\bm{h}_{i2}}},\dots,\frac{\bm{h}_{iM}}{\sqrt{\bm{h}_{iM}^\her\mathbf{G}_i^\subn\bm{h}_{iM}}}\right),\label{eq:GEP_wn}
\end{align}
where $\bm{h}_{i2},\dots,\bm{h}_{iM}\in\mathbb{C}^M$ are the other generalized eigenvectors of (\ref{eq:GEP}).
By using (\ref{eq:GEP_wt}) and (\ref{eq:GEP_wn}), we can update all $\mathbf{W}_i$ columns simultaneously. 

We can derive the update rules for the other parameters $t_{ik}^\subtn$, $v_{kj}^\subtn$, and $\lambda_i^\subn$ in the same manner as in FastMNMF~\cite{Sekiguchi2020} by using the majorization-minimization (MM) algorithm~\cite{MM}:
\begin{align}
	t_{ik}^\subt &\leftarrow t_{ik}^\subt\sqrt{\frac{\sum_j\frac{|y_{ij1}|^2}{(r_{ij}^\subt + r_{ij}^\subn\lambda_i^\subn)^2}v_{kj}^\subt}{\sum_j\frac{1}{r_{ij}^\subt + r_{ij}^\subn\lambda_i^\subn}v_{kj}^\subt}},\label{eq:updatett}\\
	v_{kj}^\subt &\leftarrow v_{kj}^\subt\sqrt{\frac{\sum_i\frac{|y_{ij1}|^2}{(r_{ij}^\subt + r_{ij}^\subn\lambda_i^\subn)^2}t_{ik}^\subt}{\sum_i\frac{1}{r_{ij}^\subt + r_{ij}^\subn\lambda_i^\subn}t_{ik}^\subt}},\label{eq:updatevt}\\
	t_{ik}^\subn &\leftarrow t_{ik}^\subn\sqrt{\frac{\sum_j\left(\frac{\lambda_i|y_{ij1}|^2}{(r_{ij}^\subt + r_{ij}^\subn\lambda_i^\subn)^2} + \frac{\sum_{n=2}^M|y_{ijn}|^2}{(r_{ij}^\subn)^2} \right)v_{kj}^\subn}{\sum_j\left(\frac{\lambda_i}{r_{ij}^\subt + r_{ij}^\subn\lambda_i^\subn} + \frac{M-1}{r_{ij}^\subn}\right)v_{kj}^\subn}},\label{eq:updatetn}\\
	v_{kj}^\subn &\leftarrow v_{kj}^\subn\sqrt{\frac{\sum_i\left(\frac{\lambda_i|y_{ij1}|^2}{(r_{ij}^\subt + r_{ij}^\subn\lambda_i^\subn)^2} + \frac{\sum_{n=2}^M|y_{ijn}|^2}{(r_{ij}^\subn)^2} \right)t_{ik}^\subn}{\sum_i\left(\frac{\lambda_i}{r_{ij}^\subt + r_{ij}^\subn\lambda_i^\subn} + \frac{M-1}{r_{ij}^\subn}\right)t_{ik}^\subn}},\\
	\lambda_i^\subn &\leftarrow \lambda_i^\subn\sqrt{\frac{\sum_j\frac{r_{ij}^\subn|y_{ij1}|^2}{(r_{ij}^\subt + r_{ij}^\subn\lambda_i^\subn)^2}  }{\sum_j\frac{r_{ij}^\subn}{r_{ij}^\subt + r_{ij}^\subn\lambda_i^\subn}}}.\label{eq:updatelambda}
\end{align}

We obtain the extracted target source signal $\hat{\bm{s}}_{ij}$ by using the following MWF~\cite{JD}:
    \begin{align}
        \label{eq:WienerFilter}
       \hat{\bm{s}}_{ij} = 
            \underbrace{
                ~~ \bm{a}_i^\subt  ~~
                \vphantom{\textstyle \frac{r_{ij}^\subt}{r_{ij}^\subt + r_{ij}^\subn\lambda_i^\subn}}
            }_{\text{projection back\cite{MURATA20011}}} 
            \underbrace{
                \textstyle
                \frac{r_{ij}^\subt}{r_{ij}^\subt + r_{ij}^\subn\lambda_i^\subn} }_{\text{postfiltering}}
            \underbrace{
                (\bm{w}_i^\subt)^\her\bm{x}_{ij}
                \vphantom{\textstyle \frac{r_{ij}^\subt}{r_{ij}^\subt + r_{ij}^\subn\lambda_i^\subn}}
            }_{\text{linear filtering}},
    \end{align}
where $\bm{a}_i^\subt = (\mathbf{W}_i^\her)^{-1}\bm{e}_1$.
\subsection{Relationship with other methods}\label{sec:relation}
In this section, we describe the relationship between NoisyILRMA and other methods. 
When we assume the two-source situation (the target source and noise), RC-FastMNMF becomes equivalent to NoisyILRMA in terms of modeling. From (\ref{eq:GenRC}) and (\ref{eq:JD}), 
$
	\tilde{y}_{ijm}\sim \cG{0}{\sum_{\tilde{n}=1}^{\tilde{N}} r_{ij\tilde{n}}\lambda_{im\tilde{n}}}\label{eq:y_RC}
$
holds, where $\tilde{\bm{y}}_{ij} =(\tilde{y}_{ij1},\dots,\tilde{y}_{ijM})^\trp= \tilde{\mathbf{W}}_i^\her\bm{x}_{ij}$ is the decorrelated signal.
By substituting $\tilde{N} = 2$, $\lambda_{i11} = 1$, $\lambda_{im1} =0$ $(m\in\{2,\dots,M\})$, $\lambda_{i12} = \lambda_i^\subn$, and $\lambda_{im2}=1$ $(m\in\{2,\dots,M\})$, we can confirm that  $\tilde{\bm{y}}_{ij}$ satisfies (\ref{eq:RC_yt})--(\ref{eq:RC_yn}) when we assume $\tilde{\mathbf{W}}_i = \mathbf{W}_i$.
One major advantage of NoisyILRMA over RC-FastMNMF is that by setting $\tilde{N} = 2$,
we can use a fast algorithm in which all $\mathbf{W}_i$ columns are updated simultaneously, whereas each $\tilde{\mathbf{W}} _i$ column is alternately updated in RC-FastMNMF. This makes NoisyILRMA considerably faster than RC-FastMNMF.

OverILRMA~\cite{OverILRMA} is a method closely related to NoisyILRMA. The main difference is that $r_{ij}^\subn$ is a time-invariant parameter in OverILRMA, while $r_{ij}^\subn$ is modeled as a time-variant parameter using the NMF model in NoisyILRMA. There are two advantages of time-variant modeling. Firstly, we can model time-variant diffuse noise. 
Secondly, the noise component in $y_{ij1}$ can be estimated using the time synchronization with the other separated signals in time-variant modeling, whereas it is estimated independently of the other separated signals in time-invariant modeling.
Because of these advantages of time-variant modeling, NoisyILRMA is expected to achieve higher performance than OverILRMA.
\subsection{Source model switching in NoisyILRMA}
In this section, we propose a source model switching mechanism. In this method, we first use NoisyILRMA for several iterations to obtain $\mathbf{W}_i$ accurately by the NMF model.
After that, we fix $\mathbf{W}_i$ and switch to the same source model as in RCSCME, i.e., $r_{ij}^\subt$ and $r_{ij}^\subn$ are the unconstrained parameters with the inverse-gamma prior distribution in (\ref{eq:RCSCME_prior}). 
This enables a finer estimation of $r_{ij}^\subtn$ and we expect higher BSE performance using MWF (\ref{eq:WienerFilter}).

We can derive the update rules based on maximum a posteriori estimation in a manner similar to that in RCSCME~\cite{Kubo2020} by using the MM algorithm as follows (the update rule for $\lambda_i^{\subn}$ is the same form as \eqref{eq:updatelambda}):
\begin{align}
	r_{ij}^\subt &\leftarrow r_{ij}^\subt\sqrt{\frac{\frac{|y_{ij1}|^2}{(r_{ij}^\subt + r_{ij}^\subn\lambda_i^\subn)^2}+ \frac{\beta}{(r_{ij}^\subt)^2}}{\frac{1}{r_{ij}^\subt + r_{ij}^\subn\lambda_i^\subn} + \frac{\alpha+1}{r_{ij}^\subt}}},\label{eq:updatert_RCSCME}\\
	r_{ij}^\subn &\leftarrow r_{ij}^\subn\sqrt{\frac{\frac{\lambda_i|y_{ij1}|^2}{(r_{ij}^\subt + r_{ij}^\subn\lambda_i^\subn)^2} + \frac{\sum_{n=2}^M|y_{ijn}|^2}{(r_{ij}^\subn)^2} }{\frac{\lambda_i}{r_{ij}^\subt + r_{ij}^\subn\lambda_i^\subn} + \frac{M-1}{r_{ij}^\subn}}}.\label{eq:updatern_RCSCME}
\end{align}
\section{Experimental analysis}\label{sec:exp}
\subsection{Experimental conditions}
\begin{figure}[tbp]
\centering
\centerline{\includegraphics[width=7.3cm]{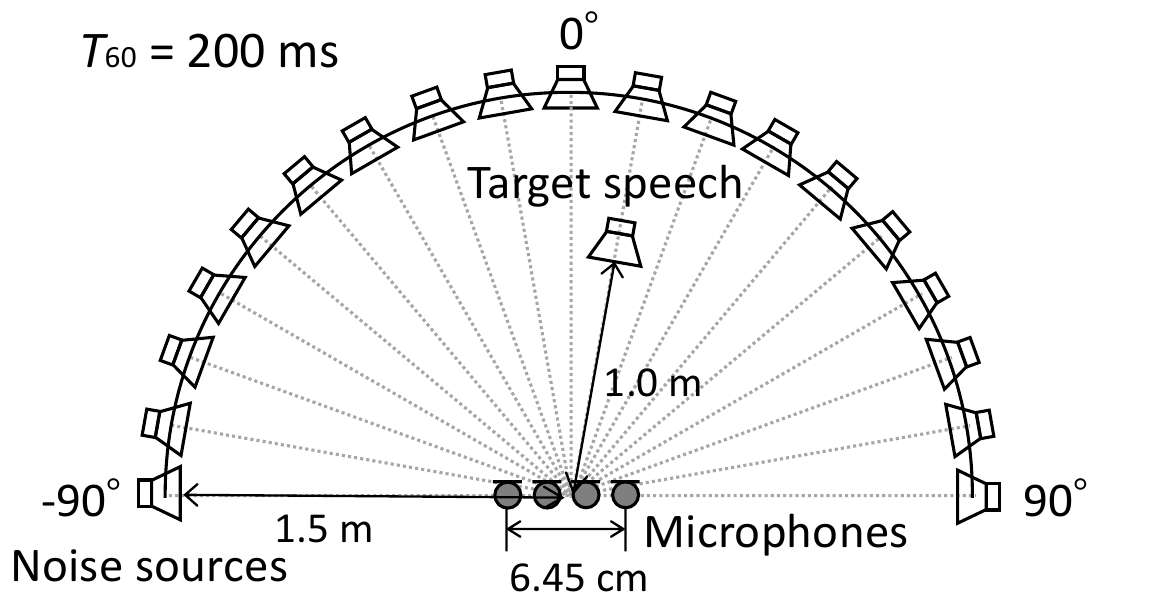}}
\caption{Recording conditions of impulse responses.}
\label{fig1}
\end{figure}

We conducted a BSE experiment using simulated mixtures of a target source and diffuse noise in $M = 4$. For the target source, we used four different directions: 0, 10, 20, and 30 degrees clockwise from the normal to the microphone array. As the target source signals, we used six speech signals from JNAS~\cite{Katunobu1999}. For diffuse noise, we used four types of noise: babble, cafe, station, and traffic. We simulated diffuse noise by different signals arriving from 19 directions. The babble noise was prepared from the speech of 19 speakers in JNAS. For cafe, station, and traffic noises, we obtained noise signals from the DEMAND~\cite{Thiemann2013} dataset and split them into 19 signals. Each signal was convoluted with the impulse response shown in Fig.~\ref{fig1}. The signal length was 8.8~s and the sampling frequency was 16~kHz. The input SNR was set to 0~dB. For STFT, a 64-ms-long Hamming window was used, and the frameshift was 32~ms.  The source-to-distortion ratio (SDR)~\cite{Vicent2006} improvement was used to evaluate the BSE performance.

The compared methods were RC-FastMNMF~\cite{Sekiguchi2020}, OverILRMA~\cite{OverILRMA}, RCSCME~\cite{Kubo2020}, proposed NoisyILRMA, and proposed NoisyILRMA with switching. In RCSCME, we attempted 20 and 50 iterations for the preprocessing ILRMA, which are labeled ``ILRMA 20'' and ``ILRMA 50'', respectively. For all the methods using the NMF model, the numbers of NMF bases for the target sound and noise were set to three. All NMF variables were initialized by uniform random numbers on (0,1) intervals. $\mathbf{W}_i$ was initialized as $\bm{w}_{im} = \bm{u}_{im}/\sqrt{d_{im}}$, where $d_{im} > 0$ and $\bm{u}_{im}\in\mathbb{C}^M$ are the eigenvalue and eigenvector of $\sum_j\bm{x}_{ij}\bm{x}_{ij}^\her/J$, respectively, and $d_{i1}$ is the largest eigenvalue. $\tilde{\mathbf{W}}_i$ was initialized in the same manner as $\mathbf{W}_i$. $\lambda_{im\tilde{n}}$, $\lambda_i^\subn$, and $\mu_i$ were initialized as one. In RCSCME, $\bm{b}_i$ was fixed to $\bm{a}'_i$ to achieve good performance. In OverILRMA and NoisyILRMA, the proportion between the number of $\mathbf{W}_i$  updates and the others was experimentally determined to be 1:10. In NoisyILRMA with switching, the number of iterations before switching was experimentally determined to be four. All methods were implemented in MATLAB (R2022a) and the calculation was performed on Intel Core i9-12900K.
\begin{figure}[tbp]
\centering
\centerline{\includegraphics[width=9cm]{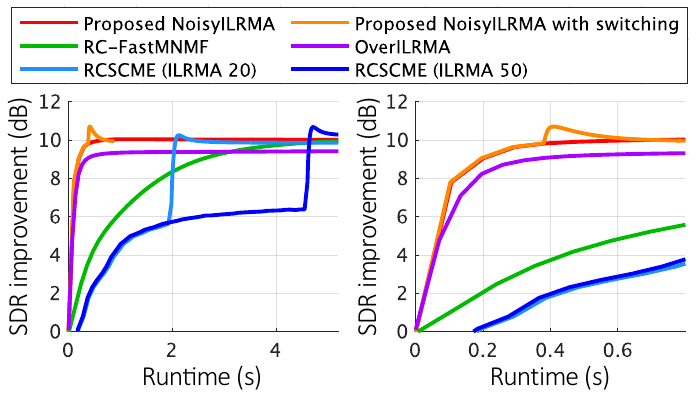}}
\caption{(a) SDR improvement behaviors with respect to runtime averaged over 960 trials. (b) Enlarged view of (a). }
\label{fig:SDRi_main}
\end{figure}
\subsection{Experimental results}
SDR improvement versus runtime is shown in Fig.~\ref{fig:SDRi_main}.
The runtime and SDR improvement were averaged over 960 trials (10 random initializations, four target speech directions, six speech signals, and four noise types).
Fig.~\ref{fig:SDRi_main} shows that the proposed NoisyILRMA was the fastest among all the methods while maintaining the convergence performance of RC-FastMNMF. In addition, NoisyILRMA achieved approximately 1~dB higher performance than OverILRMA. 
This may be due to the time-variant noise modeling in NoisyLRMA. Fig.~\ref{fig:SDRi_main}  also shows that RCSCME achieves good maximum performance with 50 ILRMA iterations and that NoisyILRMA with switching achieved comparable performance about 10 times faster.
\section{Conclusion}
 In this paper, we proposed NoisyILRMA, a diffuse-noise-aware ILRMA, which can be optimized with the same fast algorithm as proposed in FIVE. We also proposed a switching mechanism of NoisyILRMA to further improve the BSE performance. The experimental result showed that NoisyILRMA ran faster than the conventional methods while maintaining the BSE performance. We also confirmed that the switching mechanism improves the BSE performance of NoisyILRMA to become comparable to that of RCSCME at an approximately 10 times faster speed. 

\bibliographystyle{IEEEtran}

\end{document}